\documentclass[prl,showpacs,showkeys,twocolumn,floatfix]{revtex4}
\usepackage{latexsym,epsf,rotating}
\usepackage[dvips]{epsfig}

\begin{document}

\date{\today}
\title{Physically inspired analysis of prime number constellations}
\author{P.F.~Kelly}
\email{patrick.kelly@ndsu.nodak.edu}

\author{Terry~Pilling}
\email{terry@mailaps.org}
\affiliation{Department of Physics, North Dakota State University, 
Fargo, ND, 58105-5566}


\begin{abstract}
We adopt a physically motivated empirical approach to the characterisation of
the distributions of twin and triplet primes within the set of primes, 
rather than in the set of all natural numbers.
Remarkably, the occurrences of twins or triplets in any finite sequence 
of primes are like fixed-probability random events.
The respective probabilities are not constant, but instead depend on the 
length of the sequence in ways that we have been able to parameterise.
For twins the ``decay constant'' decreases as the reciprocal of the
logarithm of the length of the sequence, whereas for triplets the
falloff is faster: decreasing as the square of the reciprocal of the 
logarithm of the number of primes.
The manner of the decrease is consistent with the Hardy--Littlewood 
Conjectures, developed using purely number theoretic tools of analysis.
\end{abstract}

\maketitle

\section{Introduction}

Recently we discovered a novel approach to the characterisation of the 
distribution of twin primes and realised some of its 
consequences~\cite{random,returnof,revengeof}.
Our results have been confirmed and shown to be consistent over a
broader range by Wolf~\cite{newwolf}.
Twins are pairs of primes $\{ p, p+2 \}$ whose arithmetic separation
is minimised, {\it i.e.,} they consist of consecutive odd natural numbers.
The {\it Twin Prime Conjecture} posits that there are an infinite
number of twins~\cite{hardy}.
Long ago, Hardy and Littlewood applied sieve arguments to establish
relations which describe the behaviour of $\pi_{2}(N)$, the number
of twins with constituents less than natural number $N$.
They obtained:
\begin{equation}
\lim_{N \rightarrow \infty} 
\frac{\pi_{2}(N)}{ \int_{2}^{N} \frac{1}{\log( x )^2}\, dx} = 2 \, C_2 \, ,
\label{HL2}
\end{equation}
where the so-called twin prime constant, 
$C_2 \simeq 0.66016\ldots$ is calculable to much greater accuracy than
is quoted here~\cite{hardy}.

Prime triplets comprise the next level of allowed structure in the sequence
of prime numbers.
Again the arithmetic difference between the first constituent and the last
is minimised.
This minimum value is not $4$ as one might naively suppose because any 
set of odd numbers of the form $\{ n , n+2 , n+4 \}$ has an element which
is equivalent to zero(modulo 3).
Thus the minimum arithmetic difference is $6$, and hence there are two
``flavours'' of triplets:  those of the form $\{ p , p+2 , p+6 \}$,
which we call the twin-outlier ({\sc to}) type and those of the form 
$\{ p , p+4 , p+6 \}$ which we shall denote outlier-twin ({\sc ot}).

Hardy and Littlewood applied their analysis to both flavours of triplets
and determined that
\begin{equation}
\lim_{N \rightarrow \infty} 
\frac{\pi_{3}(\mbox{{\sc to}},N)}{ \int_{2}^{N} \frac{1}{\log( x )^3}\, dx} = 
\lim_{N \rightarrow \infty} 
\frac{\pi_{3}(\mbox{{\sc ot}},N)}{ \int_{2}^{N} \frac{1}{\log( x )^3}\, dx} =
C_{3} \, ,
\label{HL3}
\end{equation}
where the triplet prime constant, $C_{3} \simeq 2.8582\ldots$ is also 
known to great accuracy.
The analytic tools used to construct these results are essentially
insensitive to the flavour of the triplets and so the limits 
for {\sc to} and {\sc ot} must agree~\cite{hardy}.

There exist analogous relations for higher order constellations of 
prime numbers.
To place the formulae of the Hardy-Littlewood Conjectures in perspective,
it is useful to express the Prime Number Theorem -- that there exists
an infinite number of prime numbers -- in the following, stronger, form:
\begin{equation}
\pi_{1}(N) \sim \int_2^N \frac{1}{\ln(x)} dx\, .
\label{pi1}
\end{equation}

More recently (with the advent of electronic computing),
a number of investigators have studied in detail the actual 
distributions of primes and prime constellations.
Particular attention has been paid to enumeration of twins and 
explicit determination of $\pi_{2}(N)$ for large values of $N$, 
by Nicely~\cite{nicelydata}, and others~\cite{rensselaer,richstein}.
In many of these instances, the search for twins is a beneficient 
application of research in decentralised computing.
Other analyses have been concerned with the details of the distribution
of twins~\cite{brent1,brent2,wolf1}, and the existence and size 
of ``gaps'' in the sequence of twins~\cite{nicely1,odlyzko,wolf2}.
In all cases, to the best of our knowledge, these attempts situate the 
twins (and higher constellations) within the set of natural numbers.

There are three essential constituents of our new models for the 
distributions of twins and triplets.
First and foremost is that the distribution of twins and triplets are
viewed in the context of the sequence of primes, not the natural numbers.
Second is that for a prime sequence of length $\pi_1$, twins and triplets
occur in the manner of random fixed-probability events.
The third part of each model is that the fixed value of the probability
depends on $\pi_1$, the length of the sequence, in a fairly simple manner.

\section{Method and Results}  

We generated prime numbers in sequence, \textit{viz,}
$\ P_1 = 2, P_2 = 3, P_3 = 5, P_4 = 7, P_5 = 11 \ldots $,
and within this sequence counted \textit{prime separations} defined 
as the number of ``singleton'' primes 
occuring between adjacent pairs of twins, or triplets of a specified flavour.
There is an irregularity with our definition of separation
for the first few primes: $2\  (3\ 5) (5\ 7)$,
where a pair of twins is overlapping, yielding an anomalous
prime separation of $-1$.
Fortunately, as explained above such overlapping twins
do not ever recur and we chose to begin our sequences with $P_3 = 5$.

Note that there are many twins with prime separation equal to zero,
for example,
$(5\ 7) (11\ 13)$, or $(137\ 139) (149\ 151)$.
Similarly, there are many triplets which overlap, {\it i.e.,}
$(5\ 7\ 11) (11\ 13\ 17)$, or $(97\ 101\ 103) (103\ 107\ 109)$, 
and thus are assigned prime separation $-1$.
Incidently, all of the prime {\sc tt}-quadruplets of the form
$\left( P, P+2, P+6, P+8 \right)$
are comprised of a pair of twins with zero prime separation.
Also they may be viewed as an overlapping {\sc to}-{\sc ot} flavour 
combination of triplets.
Whereas {\sc oto}-quadruplets $\left( P, P+4, P+6, P+10 \right)$
possess a single twin shared between an overlapping {\sc ot}-{\sc to}
pair of triplets.
Clearly, $k$-tuplets, $k>2$, or higher-order ``prime constellations''
are composed of smaller, more primitive, units.

Concretely, so as to make our methods perfectly clear, consider the set of
prime numbers greater than or equal to $5$ and less than $100$.
In this range there are seven twins,
\[
\left\{ (5\ 7) (11\ 13) (17\ 19) (29\ 31) (41\ 43) (59\ 61) 
(71\ 73) \right\} ,
\]
and hence six separations.
Two of these happen to be 0, three are 1, and one is 2, so the 
relative frequencies for separations $s = 0 , 1 , 2$ are
$\frac{1}{3} , \frac{1}{2}$, and $\frac{1}{6}$, respectively.
In the set of four {\sc to}-triplets between 5 and 100,
\[
\left\{ (5\ 7\ 11) (11\ 13\ 17) (17\ 19\ 23) (41\ 43\ 47) \right\} ,
\]
there are three separations.
Two are -1, one is 3, so the relative frequencies are
$\frac{2}{3}, \frac{1}{3}$ for $s = -1, 3$.
The four {\sc ot}-triplets less than 100,
\[
\left\{ (7\ 11\ 13) (13\ 17\ 19) (37\ 41\ 43) (67\ 71\ 73) \right\} ,
\]
have three distinct separations.
The relative frequencies are $\frac{1}{3}$ for each of $s = -1, 3, 4$.

\begin{figure}[tbh]
\begin{center}
\leavevmode
\epsfxsize=3.0in
\epsfysize=3.0in
\epsfbox{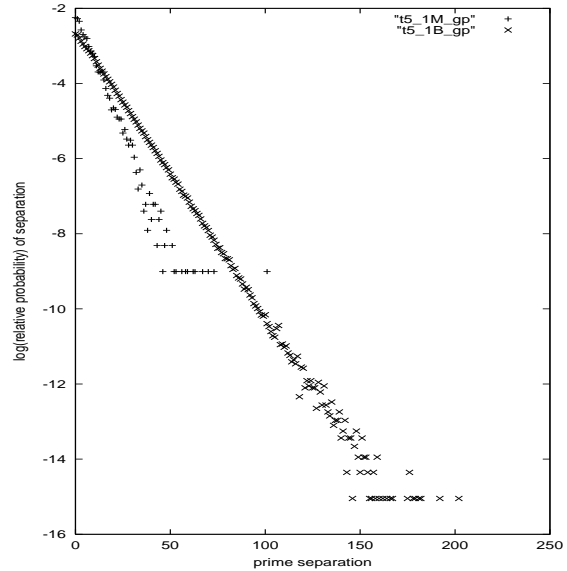}
\end{center}
\caption{Distribution of twin prime separations for $N = 1 \times 10^6$
and $N = 1 \times 10^9$. Note the (approximately) linear behaviour, and 
the differing slopes.}
\label{graph1}
\end{figure}

\begin{figure}[tbh]
\begin{center}
\leavevmode
\epsfxsize=3.0in
\epsfysize=3.0in
\epsfbox{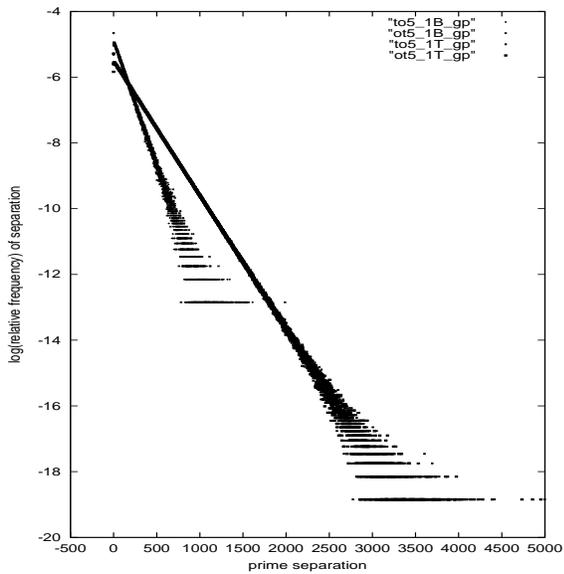}
\end{center}
\caption{Distribution of triplet prime separations for $N = 1 \times 10^9$
and $N = 1 \times 10^{12}$. The distributions for the two flavours overlap
to such an extent that they are virtually indistinguishable.
Note the (approximately) linear behaviour, 
the differing slopes, and the difference in scale from the twins case.}
\label{graph2}
\end{figure}

All separations between pairs of twins and {\sc to}- {\sc ot}-triplets
up to $N$ were computed and tabulated.
[We chose values of $N$ which ranged from less than $10^5$,
the limit of poor statistics, to $6 \times 10^{12}$, the limit of 
our patience.]
We also obtained $\pi_{1}(N)$ for each of these ranges and the count of
the number of singleton primes that occur above the last tuple in the 
range.
Taking the logarithm of the relative frequency of occurrence of each 
separation in the sequence of primes to $N$ and plotting it \textit{vs.}
the separation yields a surprisingly simple \textit{linear} relation
as illustrated in Figures~\ref{graph1} and~\ref{graph2}.

From the figures, we see that the twins and triplets are occurring 
in a fixed-probability \textit{random} manner, but that the probability
diminishes as the length of the sequence of primes increases.
By way of analogy, consider radioactive decay.
The likelihood of an atom decaying in any short time interval is constant
for a particular substance, and thus the probability that the next decay
observed in a sample occurs at time $t$ is proportional to $\exp(-\gamma t )$,
where $\gamma$, the decay rate, is a property of the species of atom.
The values of the best-fit slopes that we measure from figures like
those above determine ``decay constants'' for the twins and triplet flavours.
We will also find it useful to think of the \textit{mean separation},
the expected number of singleton primes appearing between adjacent 
$k$-tuplets, corresponding to the \textit{mean lifetime} of the radioactive
substance.
As expected from our analogy, $\bar{s} = 1\big/\mbox{slope}$.
In our reformulation of our twin model~\cite{revengeof}, and our analysis
of triplets~\cite{pktpinprep}, in terms of the counts $\pi_{1}(N)$ and
$\pi_{2,3}(N)$ alone, we cast our analysis in terms of $\bar{s}$.

The linear fits that we employed were constrained to ensure that the 
relative frequencies are properly normalised.
With negligible deleterious effects we treat $s$ as a continuous 
variable and integrate over all possible separations:  from the 
minimum value possible to infinity.
For twins, we must have
\begin{equation}
+ \mbox{(intercept)} \equiv \ln( -\mbox{(slope)})\, , \
f(s) = - m s + \ln(m)\, ,
\end{equation}
since the minimum separation is $0$.
For triplets, we have a similar condition
\begin{equation}
f(s) = - m (s + 1) + \ln(m)\, ,
\end{equation}
on account of the fact that the minimum separation is $-1$.

All of the separations which appeared in the 
data received equal (frequency-weighted) consideration in 
our computation of best-fit slopes.
This has a consequence insofar as the large-separation, low-frequency
events constituting the tail of the distribution reduce the magnitude
of the measured slope, (see Figures~\ref{graph1} and \ref{graph2}).
One might well be inclined to truncate the data by excising the
tails and fixing the slopes by the (more-strongly-linear) low-separation
data for each $N$.
We did not do this because it would have entailed a generally systematic
discarding of data from pairs appearing near the upper limit of the
range, and thus would nearly correspond to the slope with greater magnitude
that one would expect associated with an \textit{effective} upper
limit $N_{{\hbox{\small eff}}} < N$.
Viewed from this perspective, it is better to weigh all points equally.

Were it not the case that the magnitude of the slope diminished for 
larger values of $N$, then the Hardy--Littlewood Conjectures
would certainly be false, since constant slope would mean
that the probability of a given prime being a 
member of a constellation is a universal constant.
In this case then the numbers of twins and triplets $\pi_{2,3}(N)$ 
would just be fixed fractions of $\pi_1(N)$ 
in disagreement with Hardy--Littlewood and the empirical data.

As the fixed probabilities are seen to change with $\pi_{1}$, we model
the manner in which they vary.
In the figures below, we present the estimated slopes
(with statistical errors only, the total error is expected to be 
somewhat larger), \textit{vs.} $\log(\pi_{1}(N))$ -- with correction
for the singleton primes beyond the last tuplet -- and functions
that we believe capture very well the behaviour of the ``decay constant.''

\begin{figure}[tbh]
\begin{center}
\leavevmode
\epsfxsize=3.0in
\epsfysize=3.0in
\epsfbox{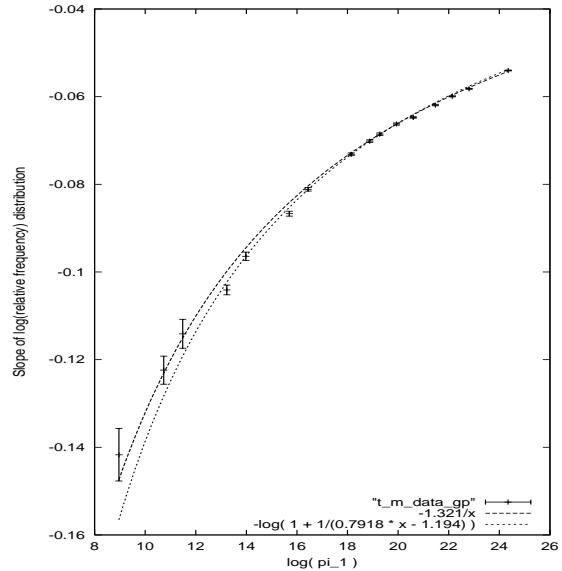}
\end{center}
\caption{Computed slopes \textit{vs.} $\log( \pi_{1}(N) )$, and 
parameterised fits for twins.
The longer dashed line is a fit to the empirical data shown, while 
the shorter dashed line arises from a analysis of prime and twin 
{\em{counts only}} up to $N \sim 10^{15}$.  }
\label{graph3}
\end{figure}

\begin{figure}[tbh]
\begin{center}
\leavevmode
\epsfxsize=3.0in
\epsfysize=3.0in
\epsfbox{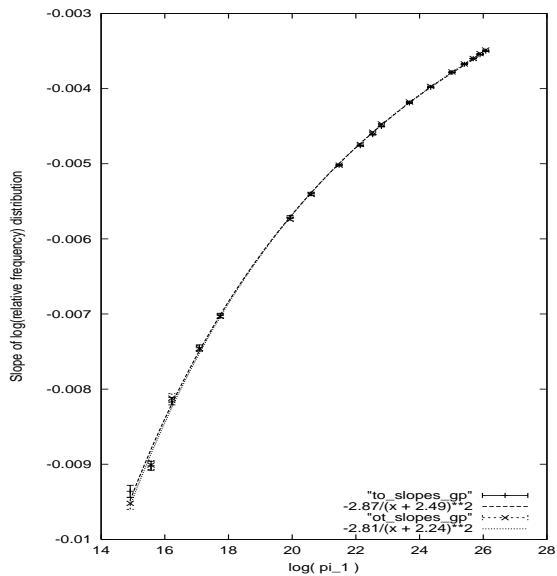}
\end{center}
\caption{Computed slopes \textit{vs.} $\log( \pi_{1}(N) )$, and 
parameterised fits for {\sc to}- and {\sc ot}-triplets.}
\label{graph4}
\end{figure}

In light of the above comments, we sketch in Figure~\ref{graph3} a plot of 
the slopes for twins \textit{versus} $\log( \pi_{1} )$.
The empirical trend seen on the graph may be well-described by
the function (remember that the error bars are understated)
\begin{equation}
- m(x) \simeq - (1.321 \pm 0.008) \big/ x \ , \quad 
\mbox{for  } x = \log( \pi_{1}(N) ) \, .
\label{t-cool-eh}
\end{equation}
We note that the factor which appears in the numerator is approximately
$-2 C_{2}$, the twin primes constant, as one is led to expect by a 
straightforward argument~\cite{random}.
Also, the asymptotic behaviour is consistent with the Twin Prime Conjecture.
On this graph we have also traced the curve which results from a 
discretised model for the distribution of twins that we 
developed~\cite{revengeof} which requires only the input of 
$\pi_{1}(N)$ and $\pi_{2}(N)$ and was fit to data~\cite{nicelydata} 
including values of $N$ up to $3 \times 10^{15}$, well beyond the reach 
of our complete analysis of the actual distribution of prime separations.

For the {\sc to}- and {\sc ot}-triplets similar results ensue.
\begin{eqnarray*}
- m_{\mbox{{\sc to}}}(x) &\simeq 
\frac{- 2.87 \pm 0.05}{ ( x + 2.49 \pm 0.18 )^2} 
\ , \cr 
- m_{\mbox{{\sc ot}}}(x) &\simeq 
\frac{- 2.81 \pm 0.04}{ ( x + 2.24 \pm 0.17 )^2} 
\ . \cr
\label{toot-cool-eh}
\end{eqnarray*}
The factors which appear in the numerators of these expressions
precisely bracket the triplet prime constant in a manner which is
consistent with expections~\cite{pktpinprep}.
The error that we quote arises purely from the fit to the data 
and is again most likely an understatement, so
we are not going to argue that the {\sc to}- and {\sc ot}-distributions
are fundamentally dissimilar.

\section{Conclusion}

We believe that we have consistently extended our construction of 
a novel characterization of the distribution of twin primes to the 
prime triplets.
The most essential feature of our approach is that we consider the 
spacings of twins and triplets among the \textit{primes} themselves, 
rather than among the natural numbers.
Secondly, we modelled the distribution empirically -- 
without preconceptions -- 
and found that the twins and triplets appear amongst the sequence 
of primes in a manner characteristic
of a completely random, fixed probability system.
Again working empirically, we were able to simply parameterise
the variation of the ``decay constant''
in terms of $\pi_1$, as suggested by our outlook.

Precise details of the triplets case will be reported upon in a 
forthcoming paper~\cite{pktpinprep}.
Future work includes extension to larger ranges of data, higher-order
constellations, examination of constellation correlations, and possible
fractal interpretations.


\begin{thebibliography}{50}
%
\bibitem{random} P.F.~Kelly and Terry Pilling, {\it Characterization of the
Distribution of Twin Primes}, \eprint{math.NT/0103191}.
\bibitem{returnof} P.F.~Kelly and Terry Pilling, {\it Implications of a
New Model of the Distribution of Twin Primes}, \eprint{math.NT/0104205}.
\bibitem{revengeof} P.F.~Kelly and Terry Pilling, {\it Discrete Reanalysis
of a New Model of the Distribution of Twin Primes}, \eprint{math.NT/0106223}.
\bibitem{newwolf} Marek Wolf, {\it Some Remarks on the Distribution of Twin Primes}, \eprint{math.NT/0105211}.
\bibitem{frenchbook} G{\' e}rald Tenenbaum and Michel Mend{\` e}s France,
{\it Les Nombres Premiers}, Presses Universitaires de France, Paris, 1997.
English edition, tr. by Philip G.~Spain,
American Mathematical Society, Providence RI, 2000.
\bibitem{hardy} {G.H. Hardy and E. M. Wright, \textit{Introduction to the Theory of Numbers}, Oxford Science Publications, (1979);
\hfill\break G. H. Hardy and J. E. Littlewood, 
Acta Mathematica, {\bf 44} 1, (1922).}
\bibitem{nicelydata} Thomas R.~Nicely, Tabulated values of $\pi_{1}(N)$ 
and $\pi_{2}(N)$ can be found at \url{http://www.trnicely.net/index.html}.
\bibitem{rensselaer} Patrick Fry, Jeffery Nesheiwat, Boleslaw K.~Szymanski,
see \url{http://www.cs.rpi.edu/research/twinp/main.html}.
\bibitem{richstein} J.~Richstein, {\it Verifying the Goldbach Conjecture
up to $10^{14}$}, to be published in Mathematics of Computation.
\bibitem{brent1} Richard P. Brent, 
Mathematics of Computation, {\bf 29} 43, (1975).
\bibitem{brent2} Richard P. Brent, 
Mathematics of Computation, {\bf 28} 315, (1974).
\bibitem{pktpinprep} P.F.~Kelly and Terry Pilling, in preparation.
\bibitem{nicely1} Thomas R. Nicely, 
Mathematics of Computation, {\bf 68}:227 1311, (1999).
\bibitem{odlyzko} A. Odlyzko, M. Rubenstein and M. Wolf, 
Experimental Mathematics, {\bf 8} 107, (1999).
\bibitem{wolf1} Marek Wolf, \textit{Unexpected Regularities in the Distribution of Prime Numbers}, w Proc. of the 8th Joint EPS -- APS Int. Conf. Physics Computing, 1996.
\bibitem{wolf2} Marek Wolf, \textit{Some conjectures on the gaps between consecutive primes}, preprint IFTUWr 894/95.
%
%
%
\end{thebibliography}
\end{document}